\documentclass{article}
\usepackage{spconf,amsmath,graphicx}
\usepackage{verbatim}
\usepackage{siunitx}
\usepackage{etoolbox}
\usepackage{tabularx}
\usepackage{makecell}
\newrobustcmd{\B}{\bfseries}
\newrobustcmd{\R}{\color{red}}

\title{EMGSE: Acoustic/EMG Fusion for Multimodal Speech Enhancement}
\name{Kuan-Chen Wang$^{1}$ \qquad Kai-Chun Liu$^{2}$\qquad Hsin-Min Wang$^{3}$ \qquad Yu Tsao$^{2}$}
\address{
    $^{1}$Graduate Institute of Communication Engineering, National Taiwan University, Taiwan\\
    $^{2}$Research Center for Information Technology Innovation, Academia Sinica, Taiwan\\
    $^{3}$Institute of Information Science, Academia Sinica, Taiwan\\
   \normalsize{r10942076@ntu.edu.tw, t22302856@citi.sinica.edu.tw, whmat@iis.sinica.edu.tw, yu.tsao@citi.sinica.edu.tw}
}
\begin{document}
\ninept
\maketitle
\begin{abstract}
Multimodal learning has been proven to be an effective method to improve speech enhancement (SE) performance, especially in challenging situations such as low signal-to-noise ratios, speech noise, or unseen noise types. In previous studies, several types of auxiliary data have been used to construct multimodal SE systems, such as lip images, electropalatography, or electromagnetic midsagittal articulography. In this paper, we propose a novel EMGSE framework for multimodal SE, which integrates audio and facial electromyography (EMG) signals. Facial EMG is a biological signal containing articulatory movement information, which can be measured in a non-invasive way. Experimental results show that the proposed EMGSE system can achieve better performance than the audio-only SE system. The benefits of fusing EMG signals with acoustic signals for SE are notable under challenging circumstances. Furthermore, this study reveals that cheek EMG is sufficient for SE.
\end{abstract}
\vspace{-1em}
\section{Introduction}
\label{sec:intro}
\vspace{-1em}
It is inevitable that speech signals will be contaminated by certain types of ambient noise in daily life. To recover clean speech signals from noisy signals, speech enhancement (SE) is applied to improve the speech quality and intelligibility. SE is critical for various speech-related applications and increases their robustness in real-world environments, such as automatic speech recognition (ASR)~\cite{pandey2021dual,kinoshita2020improving}, speaker recognition~\cite{abd2020text,shi2020robust}, hearing aids~\cite{park2020speech} and cochlear implants~\cite{goehring2017speech}. Many methods have been developed to conduct SE, including the spectral subtraction~\cite{Spectralsubtraction_boll1979spectral}, Wiener filtering~\cite{Wiener_scalart1996speech}, minimum mean square error (MMSE) estimator~\cite{MMSElog_ephraim1984speech} and Kalman filtering~\cite{Kalman_paliwal1987speech}. Recently, neural network (NN) based methods have been widely applied in this research area owing to the outstanding nonlinear mapping capability of an NN. Different types of deep learning models have been applied to SE, such as fully connected neural networks~\cite{lu2013speech,xu2014regression}, convolutional neural networks (CNNs)~\cite{fu2016snr}, a fully convolutional network (FCN)~\cite{FCN_fu2017raw, FCN_tseng2020study}, a recurrent neural network (RNN)~\cite{RNN_valentini2016investigating} and long short-term memory (LSTM) models~\cite{LSTM_weninger2015speech}. Some approaches combine multiple types of models to better extract spatial and temporal information~\cite{hsieh2020wavecrn}. 
Moreover, several advanced network structures or designs have provided further improvements in SE. Notable examples are generative adversarial networks~\cite{GAN_fu2019metricgan} and attention-based networks~\cite{peng2021attention}. These approaches are verified to outperform conventional SE algorithms in terms of their performance and robustness.  

Although NN-based methods have achieved a significant improvement and have become mainstream in SE, some challenging conditions, such as low signal-to-noise ratios (SNRs) or speech noise corruption, can still compromise the effectiveness of an audio-only SE system. To deal with such challenges, numerous studies have adopted auxiliary data in SE. These data offer information on articulatory motion in different aspects without contamination by air background noise. For instance, lip images~\cite{Lip_chuang2020improved}, bone-conducted microphone signals~\cite{yu2020time}, electropalatography (EPG)~\cite{EPG}, and electromagnetic midsagittal articulography (EMMA)~\cite{EMMA_chen2020study} have all been verified to be feasible for constructing multimodal SE systems. However, these data have certain disadvantages. For example, the quality of the lip images may be affected by background scenes and quick head movements. Some types of articulatory data (e.g., EPG or EMMA) that install sensors within the mouth may cause user discomfort. To address these disadvantages, we adopted facial electromyography (EMG) approach for the proposed SE system. Facial EMG measures the activation potentials of the human articulatory muscles by attaching single or array electrodes to specific places such as the cheek and chin. EMG sensors are more comfortable for users because they can be attached to the surface of the skin and record the signals noninvasively.

In this paper, we propose a novel multimodal SE system that fuses facial EMG and audio signals for SE, namely EMGSE. Experiments were conducted on the data of an open-access corpus, CSL-EMG\_Array~\cite{diener2020csl}. A baseline system with audio-only SE was applied for comparison with EMGSE, and the performance was evaluated using objective metrics. Experiment results show that the performance of the EMGSE system surpasses that of the audio-only SE system. Furthermore, the EMGSE is more robust under difficult conditions, including low SNRs and speech noise. Therefore, the EMGSE can be applied in scenarios where noisy speech signals are still available and clear communication is necessary for certain purposes \cite{EMMA_chen2020study}, such as security guards or field agents performing duties in the site of a famous sports event.

The remainder of this paper is organized as follows. Section 2 introduces related studies conducted in this area. Section 3 presents the architecture of the proposed EMGSE system. Section 4 presents the experimental details and results used to illustrate the performance of the proposed method. Finally, Section 5 provides some concluding remarks regarding this research.
\vspace{-1em}
\section{Related work}
\label{sec:related}
\vspace{-1em}
\subsection{Neural-network-based SE}
\label{ssec:DL}
Currently, NN-based methods dominate the research field of SE owing to the powerful nonlinear mapping capability of deep learning models. In general, these methods can be categorized into two types: masking-based SE and mapping-based SE. Masking-based SE suppresses the noise components by estimating a mask and applying it to corrupted speech~\cite{saleem2020deep}; whereas mapping-based SE directly estimates clean speech from noisy speech. According to the input and output data type of mapping-based SE, two types of SE methods can be derived: waveform-mapping- and spectral-mapping-based SE. Several multimodal SE systems have achieved  good performance when using the spectral-mapping-based method~\cite{EPG,EMMA_chen2020study}. Thus, the proposed EMGSE also applies a spectral-mapping-based SE method, and its network is constructed with  rectified linear units (RELU), fully connected (FC) layers, and bidirectional long short-term memory (BLSTM). The properties of the BLSTM model are briefly described in the next paragraph.

BLSTM has been proven to be a powerful model for sequential learning tasks. It can utilize both the past and future contexts of the data using two LSTM hidden layers. One of the layers processes the data sequence forward, and the other processes it backward. In addition, LSTM can cope with the gradient vanishing and gradient exploding problems. Owing to its capabilities and advantages, BLSTM has been applied in many speech-related applications, such as speech recognition~\cite{pandey2021dual}, speaker recognition~\cite{xue2017improving} and SE~\cite{LSTM_weninger2015speech,peng2021attention,EMMA_chen2020study,EPG}. The following equations show the operation of the LSTM network used in EMGSE:
\begin{align}
    i_t &= \sigma \left( W_{ii}x_t + b_{ii} + W_{hi} \, h_{t-1} + b_{hi} \right),\label{eq:lstm}\\
    f_t &= \sigma\left(W_{if}x_t + b_{if} + W_{hf} \, h_{t-1} + b_{hf} \right),\label{eq:1}\\
    g_t &= \tanh\left(W_{ig}x_t + b_{ig}+W_{hg} \, h_{t-1} + b_{hg} \right),\\
    o_t &= \sigma\left(W_{io}x_t + b_{io} + W_{ho} \, h_{t-1} + b_{ho} \right),\\
    c_t &= f_t \odot c_{t-1} + i_t \odot g_t, \\
    h_t &= o_t \odot \tanh(c_t),
\end{align}
where $x$ is the input; $i$, $f$, $o$, and $g$ are the input, forget, output, and cell gates, respectively; $b$ is the bias vector; $c$ is the cell state vector; and $h_t$ and $h_{t-1}$ are the hidden state vectors at times $t$ and $t-1$, respectively. In addition, $\odot$ denotes the Hadamard product (element-wise product), and $\sigma$ is the sigmoid activation function.
\vspace{-1em}
\subsection{Multimodal SE}
\label{secc:fusion_strategy}
Multimodal SE incorporates different articulatory data with noisy speech signals to improve the robustness of SE systems. Various data types have been used as auxiliary information for multimodal SE systems, including visual cues~\cite{FCN_tseng2020study,Lip_chuang2020improved}, bone-conducted microphone recorded waveforms~\cite{yu2020time}, EMMA~\cite{EMMA_chen2020study} and EPG~\cite{EPG}. Moreover, different fusion strategies were used in these studies. In \cite{EMMA_chen2020study}, unilateral concatenating exhibited the best performance, whereas in~\cite{EPG}, early fusion performed better than late fusion on SE. However, the late fusion strategy has been proven to be more effective under many different scenarios~\cite{Lip_chuang2020improved,yu2020time}. Each data type is input into the network separately and subsequently merged into a single vector after encoding. In this study, the proposed EMGSE utilizes a late fusion strategy to combine EMG signals and a noisy audio spectrogram for SE. 

\subsection{Speech-related tasks and open access corpus of facial EMG}
\label{secc:emg_application}
\begin{figure}
    \centering
    \includegraphics[width=\columnwidth]{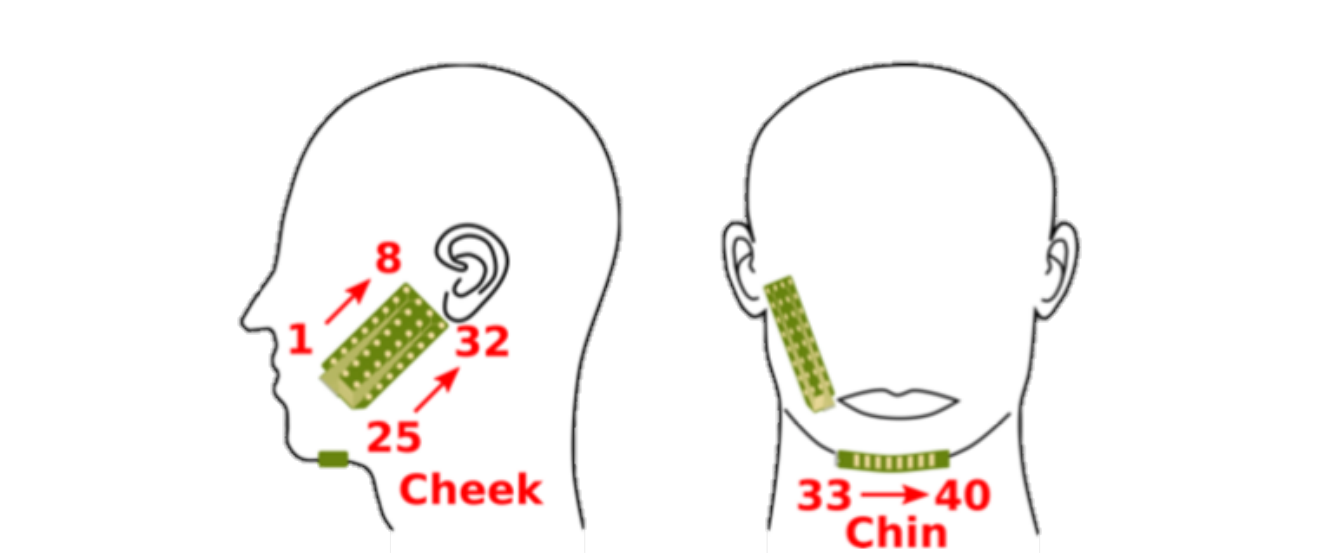}
    \caption{Placement of EMG array electrodes in CSL-EMG\_Array corpus~\cite{diener2020csl}. There are $40$ channels of EMG signals recorded during the speech.}
    \label{fig:EMG_array_position}
\end{figure}

EMG signals can indicate muscle activity. Such signals can be measured by electrodes attached to the skin without any invasive sensors or complex installations. Many studies have demonstrated the feasibility of using facial EMG in speech-related applications, such as speech recognition~\cite{jou2006towards,scheme2007myoelectric} and generation~\cite{janke2017emg,diener2018investigating}.  Thus, this study applies EMG signals to the proposed multimodal SE system as auxiliary data. The experimental data used in this study are the CSL-EMG\_Array corpus provided by Diener et al~\cite{diener2020csl}. This corpus was developed mainly for EMG-to-speech studies. In this corpus, audio and EMG signals were recorded simultaneously, and the sampling rates were 16 kHz and 2048 Hz, respectively. The EMG signals were recorded with two array electrodes placed at the cheek and chin, as shown in Fig. \ref{fig:EMG_array_position}~\cite{diener2020csl}. Furthermore, 5 cross-row channels were excluded in this study, and therefore only 35 EMG channels were used for SE. There were 12 sessions in the corpus, and each session involved 7 recording blocks to reflect the reality of the speech-conversion scenario. We only used the data of block 1 (containing audio and EMG data for 340 English utterances) in audible sessions from 8 speakers, as the main research topic focuses on SE. 

\vspace{-1em}
\section{Proposed methods}
\label{sec:proposed}
\vspace{-1em}
In this section, we describe the details of the proposed EMGSE. The feature extraction of EMG and audio is shown in Subsection 3.1. The overall network structure and fusion method of the EMGSE are illustrated in Subsection 3.2.
\vspace{-1em}
\subsection{ Feature extraction }
\label{ssec:structure}
 A feature extraction process is required to extract muscle movement information because raw EMG signals are too noisy to be used directly. Some studies have shown that the time domain (TD) feature set of EMG is suitable for speech recognition and generation~\cite{diener2020csl,janke2017emg,diener2018investigating}. Therefore, we referred to the TD15 feature set \cite{diener2020csl} for our EMGSE. Initially, high-pass and low pass third-order Butterworth filters with a cutoff frequency of 134 Hz were employed to separate the raw EMG data into high- and low-frequency parts. A Blackman window of 32 ms length and 8 ms shift was then applied to extract the TD feature set from each frame. The feature set is calculated as follows (from left to right: mean and power of low-frequency part and absolute-value mean, power and zero-crossing rate of high-frequency part):
\begin{multline}
\label{eq:TDF}
    TD(x)=[\frac{1}{n} \mathop{\sum_{k=1}^{n}}x_{low}[k],\frac{1}{n}\mathop{\sum_{k=1}^{n}}(x_{low}[k])^2,\frac{1}{n}\mathop{\sum_{k=1}^{n}}\lvert  x_{high}[k]\rvert,\\ \frac{1}{n}\mathop{\sum_{k=1}^{n}}(x_{high}[k])^2,ZCR(x_{high})],
\end{multline}
where $x$ is the EMG data of a frame, and $x_{low}$ and $x_{high}$ are the low- and high-frequency parts of $x$, respectively.

To obtain the context information, 15 frames in the past and future were stacked to form a TD15 vector (with 31 frames) of an EMG channel. Then we stack TD15 vectors from 35 EMG channels to form the input vector of EMGSE with dimensions of $35$ (channels) $\times 31$ (frames) $\times5$ (features) $ = 5,425$. Finally, we applied min-max normalization to the feature vector and made its value range from 0 to 1.

For the acoustic signal, we applied STFT to transform the noisy audio signals into the spectrogram. We also used a Blackman window of 32 ms length and 8 ms shift in STFT to match the condition of the EMG TD features. Finally, log1p and min-max normalization were applied to the spectrogram.
\vspace{-1em}
\subsection{EMGSE network structure}
\label{ssec:EMGSE_structure}
Fig. \ref{fig:EMGSE_structure} shows the overall network structure of EMGSE. Initially, the EMG and audio encoders compress both feature vectors into 100 dimensions. Both encoders are constructed through 2 FC layers (with 200 and 100 dimensions, respectively) using RELU. A dropout of 0.5 is added in all layers of EMG encoders to compensate for the relatively small EMG dataset. The SE network (SENet) is then further applied to transform the latent vector into a clean audio spectrogram. Two encoded vectors are initially fused into a latent vector with dimensions of 200 through an FC layer using RELU. The remaining parts of SENet are a BLSTM layer with 2 hidden layers (output dimensions of 500) and an output FC layer (257 dimensions) with RELU. Finally, we combine the output spectrogram with the phase component of the corresponding noisy audio and conduct an inverse STFT to reconstruct the enhanced audio waveform. This process can be expressed as follows:

\begin{figure}
    \centering
    \includegraphics[width=\columnwidth]{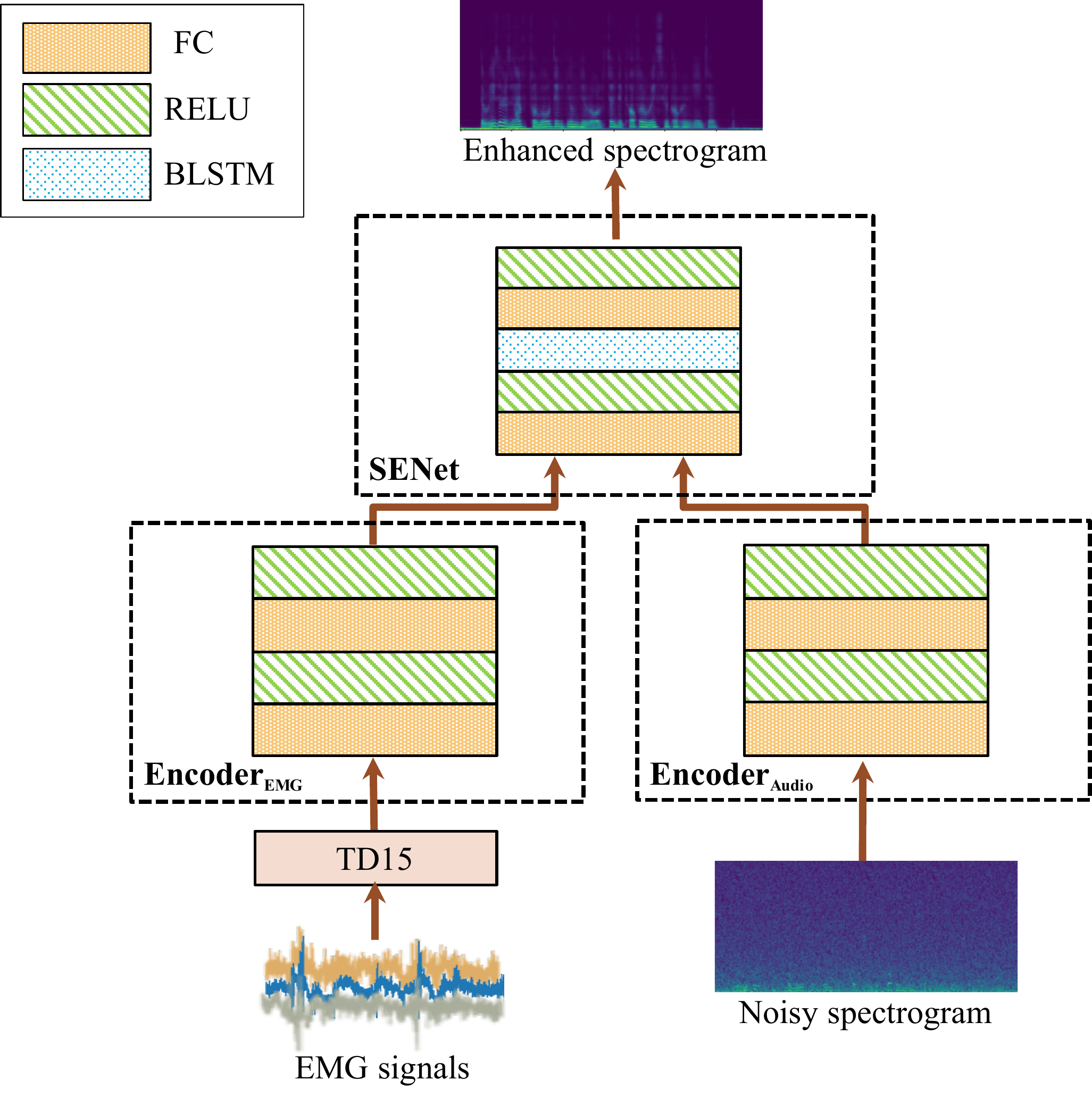}
    \caption{Architecture of EMGSE}
    \label{fig:EMGSE_structure}
\end{figure}

\vspace{-5pt}
\begin{equation}
\label{eq:enc_EMG}
    v_{EMG}[n] = \text{Encoder}_{EMG}\{x^{EMG}[n]\},
\end{equation}
\begin{equation}\label{eq:enc_Audio}
    v_{Audio}[n] = \text{Encoder}_{Audio}\{x^{Audio}[n]\},
\end{equation}
\begin{equation}\label{eq:SENet}
    z[n] = \text{SENet}\{v_{EMG}[n], v_{Audio}[n]\},
\end{equation}
where $x^{EMG}[n]$ and $x^{Audio}[n]$ are the EMG feature vector and noisy spectrogram at time n, respectively; $v_{EMG}[n]$ and $v_{Audio}[n]$ are the encoded vectors; and $z[n]$ is the enhanced spectrogram at time n.
\vspace{-1em}
\section{Experiments}
\label{sec:data}
\vspace{-1em}
\subsection{Experiment setup}
\label{ssec:setup}
As mentioned in Section 2, this study uses the corpus CSL-EMG\_Array data in block 1 from 8 speakers (340 utterances per speaker) to validate EMGSE. A total of 340 utterances were separated into 280, 20, and 40 utterances for training, validation, and testing, respectively. For the training and validation sets, we applied 100 types of nonspeech noise~\cite{100noise} to generate noisy audio data. Each utterance is corrupted with five randomly selected types of noise at five SNRs (-10, -5, 0, 5, and 10 dB). For the test set, we added 18 unseen noise types (car noise, engine noise, pink noise, white noise, two types of street noises, six kinds of background Chinese speakers, and six kinds of English speakers) to clean utterances at four SNRs (-11, -4, -1, and 4 dB) to cause mismatch conditions. We evaluated the performance of SE using two evaluation criteria: perceptual evaluation of speech quality (PESQ)
~\cite{PESQ} and short-time objective intelligibility (STOI)~\cite{STOI}. The PESQ score ranges from 0.5 to 4.5, and the STOI score ranges from 0 to 1. Higher PESQ and STOI scores represent  better speech quality and intelligibility, respectively.
\vspace{-1em}
\subsection{Implementation details}
\label{ssec:details}
The proposed EMGSE used the L1 loss and Adam optimizer to update the weights. The learning rate was set to 0.0001. To avoid an overfitting, we stopped training as the validation loss stopped dropping after 15 epochs and saved the network parameters with the least validation loss. We built a baseline system, i.e., SE with audio only (SE(A)) for comparison with EMGSE. SE(A) has an identical structure to EMGSE, except that only acoustic signals are used for the input data.
\vspace{-1em}
\subsection{Results and discussion}
\label{ssec:results}
\begin{figure}
    \centering
    \includegraphics[width=\columnwidth]{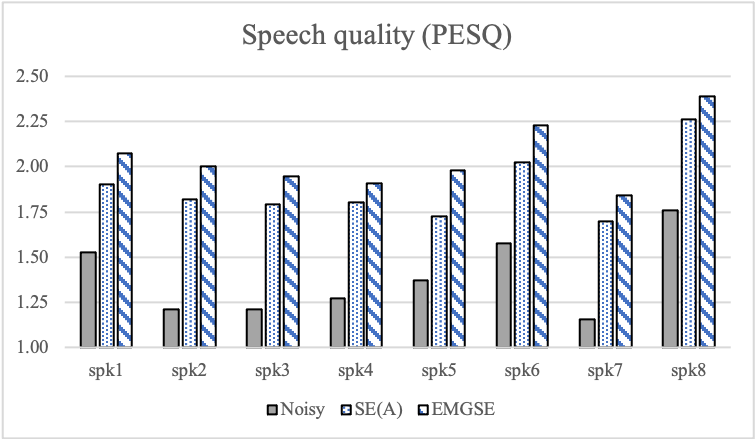}
    \includegraphics[width=\columnwidth]{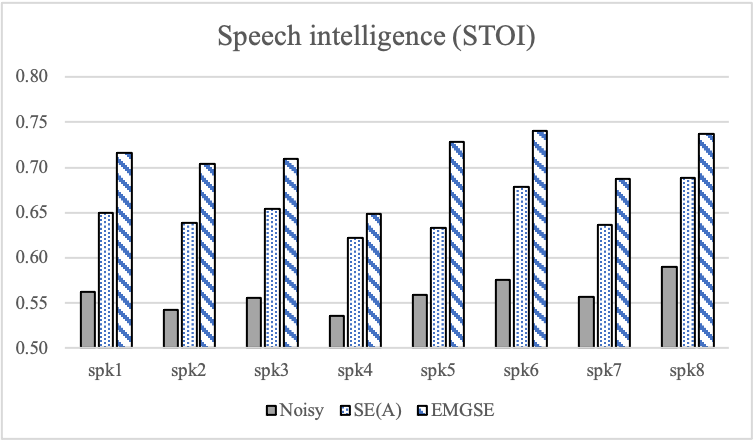}
    \caption{Performance of EMGSE and audio-only SE system (SE(A)) and noisy test data evaluated based on PESQ and STOI scores. Data on 8 speakers were tested, and the application of facial EMG improved the SE performance in every case.}
    \label{fig:Performance_8spk}
\end{figure}

Fig. \ref{fig:Performance_8spk} shows the overall performance of the two SE systems for the eight speakers. It can be seen that EMGSE achieves higher PESQ and STOI scores than SE(A) for all speakers. With facial EMG, PESQ scores can increase from 0.1 to 0.3, and STOI scores can increase by approximately 0.03 to 0.1 from case to case. To further examine the effect of  facial EMG on the SE, Table \ref{tab:evaluationSNR} presents the average performance for eight speakers under four SNRs (-11, -4 , -1, and 4 dB). The results show that EMGSE can perform particularly better than SE(A) for a low SNR (-11 dB). The PESQ and STOI scores increased by 0.225 and 0.097, respectively. Furthermore, Table \ref{tab:evaluationNoisetype} demonstrates the average performance of the two systems for different noise types. We can observe that the performance of EMGSE outperforms that of SE(A), particularly in speech noise types. Overall, the improvement with facial EMG increases as the noise conditions become tougher, and the analysis results validate that facial EMG is beneficial to multimodal SE\footnote{Noise data used in testing and the demonstration of proposed EMGSE can be viewed in https://eric-wang135.github.io/EMGSE/}.

\begin{table}
    \caption{Evaluation results on SE(A) and EMGSE under different SNRs.}
    \vspace{5pt}
    \label{tab:evaluationSNR}
    \sisetup{detect-weight,mode=text}
    \renewrobustcmd{\bfseries}{\fontseries{b}\selectfont}
    \renewrobustcmd{\boldmath}{}
    \centering 
    \begin{tabular}{c|cc|cc|cc}
            & \multicolumn{2}{c|}{\bf Noisy}
            & \multicolumn{2}{c|}{\bf SE(A)}
            & \multicolumn{2}{c}{\bf EMGSE}
            \\
    \cline{2-7}
    \cline{2-7}
            & \bf PESQ  & \bf STOI  
            & \bf PESQ  & \bf STOI   
            & \bf PESQ  & \bf STOI   
            \\
    \hline
    \bf -11dB  & 0.923 & 0.394 & 1.197 & 0.446 & \bf 1.452 & \bf 0.553 \\
    \bf -6dB  & 1.138 & 0.481 & 1.608 & 0.579 & 1.829 & 0.658 \\
    \bf -1dB & 1.448 & 0.579 & 2.03 & 0.697 & 2.207 & 0.751 \\
    \bf 4dB & 1.722 & 0.663 & 2.333 & 0.764 & 2.476 & 0.801 \\
    \hline
    \hline
    \bf Avg. & 1.308 & 0.529 & 1.792 & 0.621 & 1.991 & 0.691 \\
    \hline
    \hline
    \end{tabular}
\end{table}

\begin{table}
    \caption{Evaluation results on SE(A) and EMGSE for different noise types.}
    \vspace{5pt}
    \label{tab:evaluationNoisetype}
    \sisetup{detect-weight,mode=text}
    \renewrobustcmd{\bfseries}{\fontseries{b}\selectfont}
    \renewrobustcmd{\boldmath}{}
    \centering 
    \begin{tabular}{m{1.4cm}|m{0.75cm}m{0.75cm}|m{0.75cm}m{0.75cm}|m{0.75cm}m{0.75cm}}
            & \multicolumn{2}{c|}{\bf Noisy}
            & \multicolumn{2}{c|}{\bf SE(A)}
            & \multicolumn{2}{c}{\bf EMGSE}
            \\
    \cline{2-7}
            & \bf PESQ  & \bf STOI  
            & \bf PESQ  & \bf STOI   
            & \bf PESQ  & \bf STOI   
            \\
    \hline
    \centering \bf Chinese  &1.277 & 0.504 & 1.695 & 0.591 & \bf 1.928 & \bf 0.669  \\
    \hline
    \centering \bf English &1.290 & 0.514 & 1.677 & 0.584 & \bf 1.935 & \bf 0.676 \\
    \hline
    \centering \bf Car & 1.692 & 0.675 & 2.297 & 0.777 & 2.314 & 0.781 \\
    \hline
    \centering \bf Engine & 1.286 & 0.55 & 1.985 & 0.686 & 2.094 & 0.729 \\
    \hline
    \centering \bf Pink &1.281 & 0.552 & 1.943 & 0.658 & 2.077 & 0.713 \\
    \hline
    \centering \bf White & 1.303 & 0.585 & 2.072 & 0.691 & 2.223 & 0.737\\
    \hline
    \centering \bf Street & 1.404 & 0.564 & 2.028 & 0.704 & 2.129 & 0.743\\
    \hline
    \end{tabular}
\end{table}

\begin{figure}
    \centering
    \includegraphics[width=\columnwidth]{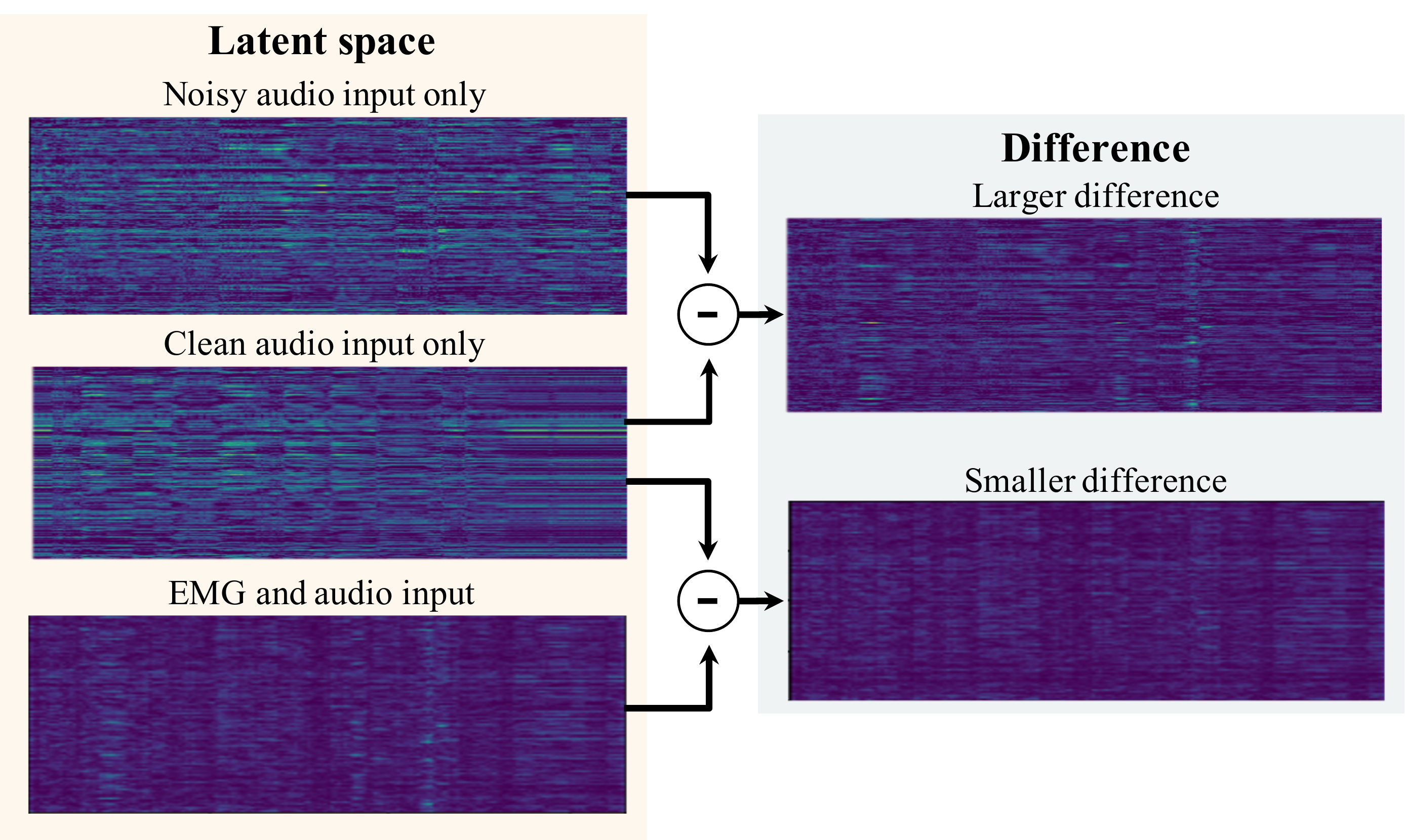}
    \caption{The latent spaces of EMGSE with different input data conditions and their difference. The latent space of EMGSE in clean-audio-input-only case can be viewed as a reference for better SE results. When facial EMG is combined with noisy audio input, the difference between the latent space and the reference minimizes, demonstrating the benefits of facial EMG.}
    \label{fig:EMGSE_latent}
\end{figure}

To study the improvement produced in the proposed system more thoroughly, we inspect the latent space and the output vector of the fusion layer using 200 dimensions in EMGSE. Fig. \ref{fig:EMGSE_latent} shows the latent spaces in audio-input-only (clean speech and noisy speech with an SNR level of -11 dB) and the EMG-audio-input case. Comparing the latent spaces of the audio-input-only cases, we can see that patterns appearing in the speech occurrence are much clearer in the clean-speech-input case. After the EMG data were applied, some patterns became more protruding. To evaluate the improved quality of the latent space, we plot the differences between the latent space of a clean-audio input and those with and without an EMG input. It can be observed that the difference minimizes after the EMG is adopted. These results again verified the effectiveness of facial EMG on the SE.

\begin{table}
    \caption{Performance of SE(A) and EMGSE with only 28 and 35 channels. A reduction in the number of channels seems to have little influence.}
    \vspace{5pt}
    \label{tab:EMGSEcheek}
    \centering 
    \begin{tabular}{c|c|c|c}
     & \bf{SE(A)} & \bf{EMGSE} & \bf \bf{EMGSE}$_\text{cheek}$\\
    \hline
    \bf{PESQ} & 1.879 & 2.046 & 2.039 \\
    \bf{STOI} & 0.65 & 0.709 & 0.711 \\
    \hline
    \end{tabular}
\end{table}

We explored the SE performance using only EMG sensors placed on the cheek (28 EMG channels). Table \ref{tab:EMGSEcheek} shows the average performance of SE(A) and EMGSE with 28 and 35 channels for the 8 speaker. To our surprise, a reduction in the number of channels had little effect on the performance. The PESQ score only decreased by approximately 0.007 and the STOI score remained almost unchanged. Therefore, it can be deduced that the cheek EMG is sufficient for SE. Reducing the number of channels produces some advantages. We can decrease the computational effort and further improve the efficiency of the SE. Moreover, users may feel more comfortable if fewer sensors are needed. These benefits increase the practicability of the EMGSE system under a real-world scenario. 
\vspace{-1em}
\section{Conclusion}
\label{sec:conclusion}
\vspace{-1em}
In this study, facial EMG was used as auxiliary data, and a non-invasive and speaker-dependent multimodal SE system, EMGSE, was proposed. To the best of our knowledge, this is the first study applying EMG to SE. Experimental results show that fusing EMG signals with acoustic signals can improve SE performance, especially under challenging circumstances, such as a low signal-to-noise ratio (SNR) and speech noise contamination. In addition, cheek EMG has been shown to be sufficient for SE, increasing the practicability of EMGSE. In the future, we plan to use EMG channel feature selection methods and EMG denoising algorithms as preprocessing to further enhance the effect of facial EMG in EMGSE. 

\bibliographystyle{IEEEbib}
{\footnotesize
\bibliography{refs}}

\end{document}